\documentclass[reprint, nofootinbib, superscriptaddress]{revtex4-1}

\usepackage{physics}
\usepackage{graphicx}
\usepackage[space]{grffile}
\usepackage{mathrsfs}
\usepackage[colorinlistoftodos]{todonotes}
\usepackage{mathtools}
\usepackage{xcolor}
\usepackage[symbol]{footmisc}
\begin{document}
\title{
Entanglement between a diamond spin qubit and a photonic time-bin qubit at telecom wavelength.}

\author{Anna Tchebotareva}
\thanks{These two authors contributed equally to this work.}
\affiliation{QuTech, Delft University of Technology, P.O. Box 5046, 2600 GA Delft, The Netherlands}
\affiliation{Netherlands Organisation for Applied Scientific Research (TNO), P.O. Box 155, 2600 AD Delft, The Netherlands}
\author{Sophie~L.~N.~Hermans}
\thanks{These two authors contributed equally to this work.}
\author{Peter C. Humphreys}
\affiliation{QuTech, Delft University of Technology, P.O. Box 5046, 2600 GA Delft, The Netherlands}
\affiliation{Kavli Institute of Nanoscience, Delft University of Technology, P.O. Box 5046, 2600 GA Delft, The Netherlands}
\author{Dirk Voigt}
\author{Peter J. Harmsma}
\author{Lun~K.~Cheng}
\author{Ad L. Verlaan}
\author{Niels Dijkhuizen}
\author{Wim de Jong}
\affiliation{QuTech, Delft University of Technology, P.O. Box 5046, 2600 GA Delft, The Netherlands}
\affiliation{Netherlands Organisation for Applied Scientific Research (TNO), P.O. Box 155, 2600 AD Delft, The Netherlands}
\author{Ana\"is Dr\'eau}
\affiliation{QuTech, Delft University of Technology, P.O. Box 5046, 2600 GA Delft, The Netherlands}
\affiliation{Kavli Institute of Nanoscience, Delft University of Technology, P.O. Box 5046, 2600 GA Delft, The Netherlands}
\affiliation{Laboratoire Charles Coulomb, Universit\'e de Montpellier and CNRS, 34095 Montpellier, France}
\author{Ronald Hanson}
\email{R.Hanson@tudelft.nl}
\affiliation{QuTech, Delft University of Technology, P.O. Box 5046, 2600 GA Delft, The Netherlands}
\affiliation{Kavli Institute of Nanoscience, Delft University of Technology, P.O. Box 5046, 2600 GA Delft, The Netherlands}

%
%
%
\begin{abstract} 
We report on the realization and verification of quantum entanglement between an NV electron spin qubit and a telecom-band photonic qubit. First we generate entanglement between the spin qubit and a 637~nm photonic time-bin qubit, followed by photonic quantum frequency conversion that transfers the entanglement to a 1588~nm photon. We characterize the resulting state by correlation measurements in different bases and find a lower bound to the Bell state fidelity of $\geq 0.77 \pm 0.03$. This result presents an important step towards extending quantum networks via optical fiber infrastructure.
\end{abstract}
\maketitle

Quantum networks connecting and entangling long-lived qubits via photonic channels~\cite{Wehner2018} may enable new experiments in quantum science as well as a range of applications such as secure information exchange between multiple nodes, distributed quantum computing, clock synchronization, and quantum sensor networks \cite{BenOr2006, Broadbent2009, LiangJiang2009, Ekert2014, Cirac1999, Gottesman2012, Nickerson2014, Komar2014, Bancal2012}. A key building block for long-distance entanglement distribution via optical fibers is the generation of entanglement between a long-lived qubit and a photonic telecom-wavelength qubit (see Fig. 1a). Such building blocks are now actively explored for various qubit platforms \cite{Dudin2010, Yamamoto2012, Eschner2018, deRiedmatten2018, Keller2018, Lanyon2019}.

The NV center in diamond is a promising candidate to act as a node in such quantum networks thanks to a combination of long spin coherence and spin-selective optical transitions that allow for high fidelity initialization and single-shot read out  \cite{Awshalom2018}. Moreover, memory qubits are provided in the form of surrounding carbon-13 nuclear spins. These have been employed for demonstrations of quantum error correction \cite{Cramer2016,Waldherr2014,Taminiau2014} and entanglement distillation \cite{Kalb2017}. Heralded entanglement between separate NV center spin qubits has been achieved by generating spin-photon entangled states followed by a joint measurement on the photons \cite{Bernien2013}.  

Extending such entanglement distribution over long distances is severely hindered by photon loss in the fibers. The wavelength at which the NV center emits resonant photons, the so-called zero-phonon-line (ZPL) at $637$~nm, exhibits high attenuation in optical glass fibers. Quantum-coherent frequency conversion to the telecom band can mitigate these losses by roughly $7$ orders of magnitude for a distance of $10$~km \cite{Miya1979, Nagayama2002} and would enable the quantum network to optimally benefit from the existing telecom fiber infrastructure. Recently, we have realized the conversion of 637~nm NV photons to $1588$~nm (in the telecom L-band) using a difference frequency generation (DFG) process and shown that the intrinsic single-photon character is maintained during this process \cite{Dreau2018}. However, for entanglement distribution an additional critical requirement is that the quantum information encoded by the photon is preserved during the frequency conversion.

Here we demonstrate entanglement between an NV center spin qubit and a time-bin encoded frequency-converted photonic qubit at telecom wavelength. The concept of our experiment is depicted in Fig.~1b. We first generate spin-photon entanglement, then convert the photonic qubit to the telecom band, and finally characterize the resulting state through spin-photon correlation measurements in different bases.

\begin{figure}[t]
	\centering
	\includegraphics[width = 0.4\textwidth]{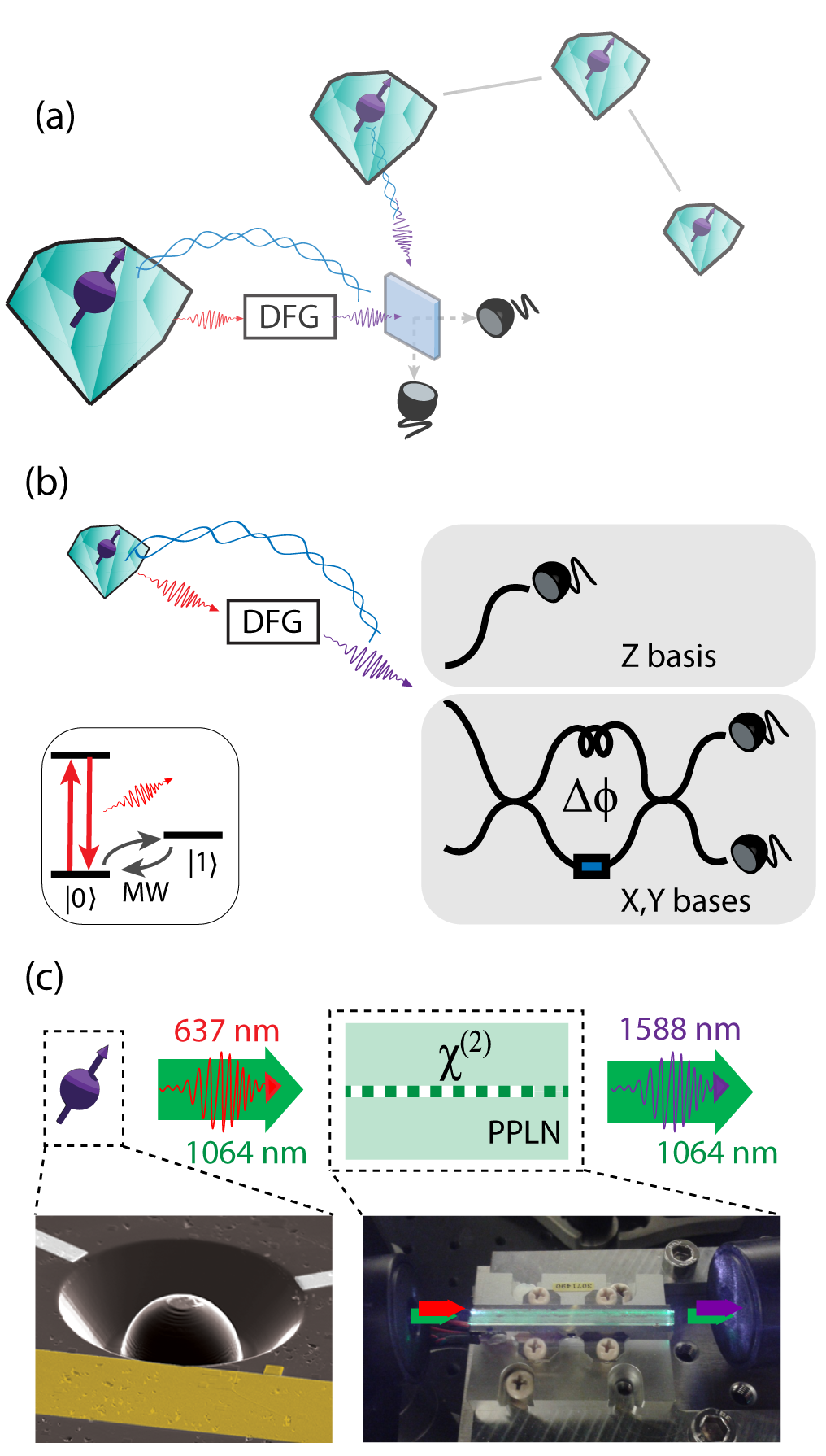}
	\caption{(a) In a long-distance quantum network, heralded entanglement between the nodes is generated by interference on a beam splitter and subsequent measurement of telecom photonic qubits. (b) Concept of the experiment. An NV spin qubit - photon qubit entangled state is generated. The photonic qubit is converted to a telecom wavelength by the difference frequency generation (DFG) process~\cite{Dreau2018}. For measuring the photonic qubit in the X and Y basis, an imbalanced interferometer is used. Inset: simplified level scheme of the NV center. (c) DFG process: a $637$~nm photon is converted to a wavelength of $1588$ nm using a nonlinear PPLN crystal. Left inset: electron microscope image of the diamond device. 
	Right inset: image of the PPLN crystal with ridge waveguides(NTT/NEL).}
\end{figure}

We use two of the NV center electron spin-$1$ sublevels as our qubit subspace. We denote the $m_s = 0$ and $m_s = -1$ ground states as $\ket{0}$ and $\ket{1}$, respectively. To generate the desired spin-photon entangled state, we first initialize the spin in $\ket{0}$ and prepare the balanced superposition $\ket{\psi} = \frac{1}{\sqrt{2}}\left(\ket{0} + \ket{1}\right)$ using a microwave $\pi/2$-pulse. Then we apply a spin-selective optical $\pi$-pulse, such that the $\ket{0}$ state will be excited, followed by photon emission (lifetime of 12~ns). Next, we flip the spin state using a microwave $\pi$-pulse and apply the optical excitation for a second time. This generates the following spin-photon entangled state:
\begin{equation}
\ket{\text{NV spin, photon}} = \frac{1}{\sqrt{2}}\ket{1,E} +\frac{1}{\sqrt{2}} \ket{0,L},
\label{eq:state}
\end{equation}
where the basis states for the photonic qubit are given by the early ($\ket{E}$) and late ($\ket{L}$) time bins, which are separated in the experiment by 190~ns, limited by the state preparation time.

Next, the photon is converted to the telecom wavelength of 1588 nm using a difference frequency generation (DFG) process, by combining it with a strong pump laser inside a periodically poled lithium niobate (PPLN) crystal waveguide (Fig. 1c) \cite{Dreau2018}. The resulting spin-telecom photon state is characterized via correlation measurements. We read out the photonic qubit in the Z basis using time-resolved detection that discriminates between the early and late time bins. 
To access other photonic qubit bases we use an imbalanced interferometer \cite{Franson1989} with a tunable phase difference $\Delta\phi$ between the two arms. For each photonic qubit basis, we read out the spin state in the basis where maximum correlation is expected. From the measured correlations in three orthogonal bases we find the fidelity to the desired maximally entangled state.

The diamond sample containing the NV center is cooled to $\approx 4$~K. The optical setup is schematically depicted in Fig.~2a. Laser light at $637$~nm is used to apply the optical $\pi$-pulses. In the photon detection path, the emitted 637~nm photons are separated from reflected excitation light using a cross-polarization configuration and time filtering. The off-resonant phonon side band emission is separated by dichroic filtering and sent to a detector (D1) for spin readout. The 637~nm photons are combined with a strong pump laser (emission wavelength of $1064$~nm) and directed into the PPLN crystal for the DFG process. Afterwards, the remaining pump laser light is filtered out by a prism, a long-pass dielectric filter and a narrow-band fiber Bragg grating. The total conversion efficiency of the DFG setup is $\eta_c\approx 17\%$ \cite{Dreau2018}. To ensure the frequency and phase stability of the converted photons, both the NV excitation laser and the pump laser are locked to an external reference cavity (Stable Laser Systems). 

Figure 2b shows the experimental sequence used in the experiments. Our protocol starts with checking whether the NV center is in the desired charge state and on resonance with the control lasers \cite{Robledo2010}. Once this test is passed, the spin-photon entangled state is generated. If a photon is detected, we read out the spin state in the appropriate basis and re-start the protocol. In case no photon is detected, we reinitialize the spin and again generate an entangled state. After 250 failed attempts to detect a photon, we re-start the protocol.

\begin{figure*}[t]
	\centering
	\includegraphics[width = 0.8\textwidth]{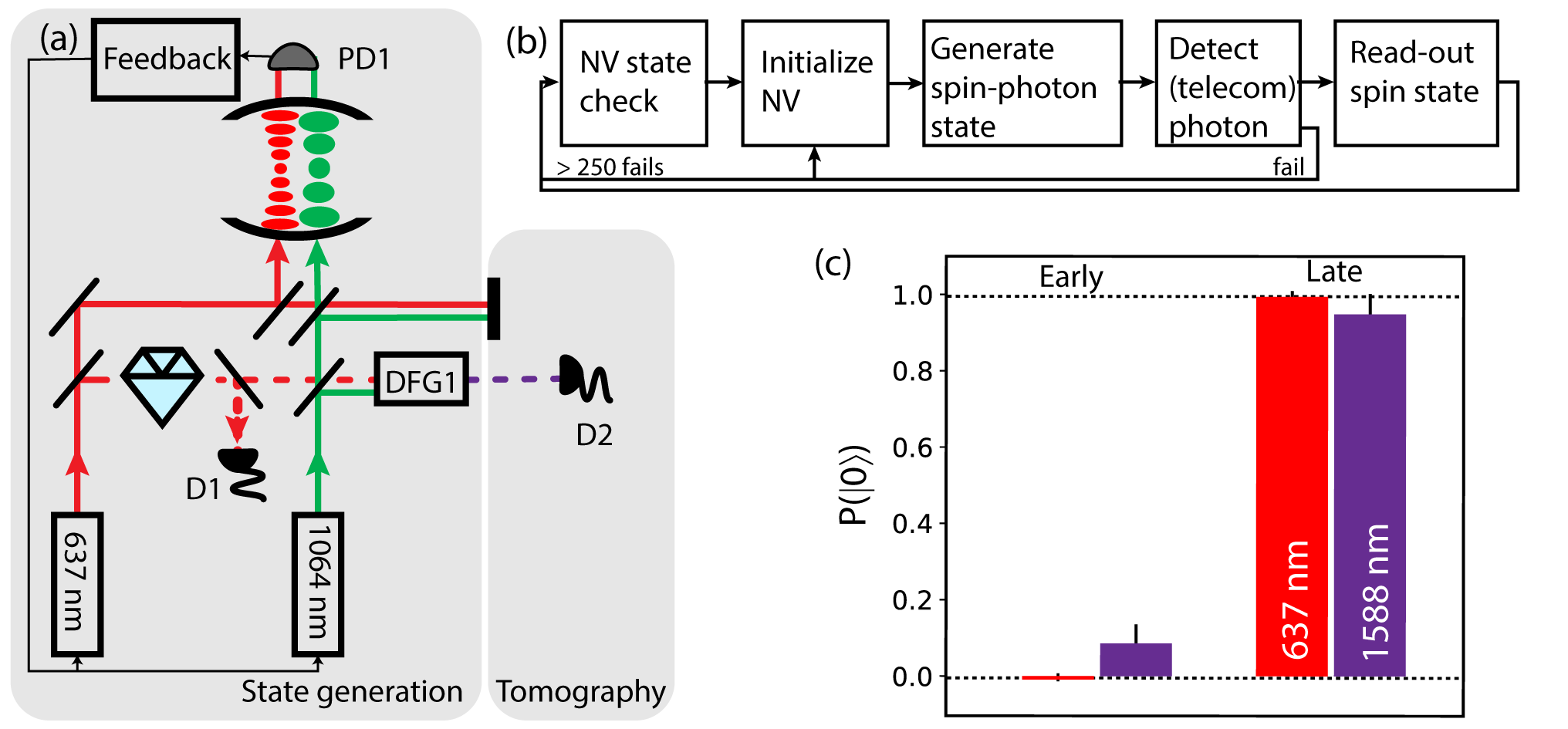}
	\caption{(a) Experimental setup for the spin-telecom photon entangled state generation. Emitted 637~nm photons are combined with the pump laser (1064~nm) in the difference frequency generation setup (DFG1). The two lasers are frequency-locked to an external reference cavity. Tomography in the Z-basis: the frequency converted photons are detected using a superconducting nanowire detector (D2) discriminating the early and late time bins. (b) Experimental protocol for generating and detecting spin-telecom photon entangled states (see main text). (c) Results for correlations measured in the Z~basis both for the red and for the frequency-converted photons at telecom wavelength. }
\end{figure*}


We first measure spin-photon correlations in the ZZ basis. To measure the photon in the Z basis, we send the frequency-converted photons directly to a superconducting nanowire detector (D2) that projects the photonic qubit in the time-bin basis, and, upon photon detection, we read out the spin qubit in the corresponding Z basis. Figure 2c shows the observed correlation data. The probability to measure the spin in $\ket{0}$ is plotted for photon detection events in the early and late time-bins. We have performed this measurement for both the 637 nm photons (red) and for the frequency-converted photons at 1588~nm (purple). For the unconverted photons we measure correlations that are perfect within measurement uncertainty (contrast of $E_{Z} = |P_E\left( \ket{0}\right) - P_L\left( \ket{0}\right)| = 0.997 \pm 0.018$). For the frequency converted photons we measure $P_E\left( \ket{0}\right) = 0.09 \pm 0.05$ for the early time bin and $P_L\left( \ket{0}\right) = 0.95 \pm 0.05$ for the late time bin, yielding a contrast of $E_{Z} = 0.86 \pm 0.07$. All data in this work are corrected for spin readout infidelity and dark counts of the detectors, both of which are determined independently.


The contrast for the telecom photons is lowered by noise coming from spontaneous parametric down converted (SPDC) photons and Raman scattering induced by the strong pump laser~\cite{Dreau2018,Fejer2010}. We characterize this noise contribution separately by blocking the incoming 637~nm path and find an expected signal to noise ratio (SNR) between $4.8$ and $7.7$. This SNR bounds the maximum observable contrast for the ZZ correlations to $0.85 \pm 0.03$, and thus fully explains our data. We use this SNR later to determine the different noise contributions for the correlation data in the other bases. Additionally, we conclude from the relative number of detection events in the early and late time bin (659 vs 642 events) that the amplitudes of the two parts of the spin-photon entangled state are well balanced.

%
%

To verify the spin-photon entanglement, we measure spin-photon correlations in two other spin-photon bases by sending the frequency-converted photons into the imbalanced fiber interferometer (see Fig.~3a). The fiber arm length difference is $\approx 40$~m, which corresponds to a photon travel time difference of $190$~ns between the two arms. In this way the early time bin taking the long arm overlaps at the second beam splitter with the late time bin taking the short arm, thus allowing us to access the phase relation between the two. To access a specific photon qubit basis, we introduce a tunable phase difference $\Delta \phi$ between the long and short arms of the interferometer. In particular, detection of a photon by the detector D3 projects the spin into the state   
\begin{equation}
	\ket{\text{NV}}_{D3} = \frac{1}{\sqrt{2}}\left( \ket{0} + e^{i\left(\Delta\phi-\frac{\pi}{4} \right)}\ket{1}\right).
\end{equation}
We use two orthogonal set points, labelled X and Y, with $\Delta\phi = \pi/4$ and $\Delta\phi = 3\pi/4$, respectively, as indicated in Fig.~3c.

A key requirement for this experiment is that the interferometer is stable with respect to the frequency of the down-converted photons; any instabilities in the interferometer will reduce the interference contrast and prevent us from accessing the true spin-photon correlations. For this reason the interferometer is thermally and vibrationally isolated. Furthermore, we split the experiment into cycles of 1 second (see Fig.~4a), of which the first 100 ms is used to actively stabilize the phase setpoint of the interferometer. Within this 100 ms, we feed metrology light into the interferometer in the reverse direction via shutter S and a circulator. This metrology light is generated by a second DFG setup, using input from the excitation and pump lasers, thus ensuring a fixed frequency relation between the metrology light and the frequency-converted photons. By comparing the light intensities on detectors PD2 and PD3 with the values corresponding to the desired $\Delta\phi$ setpoint as determined from a visibility fringe (calibrated every $100$~s), an error signal is computed and feedback is applied to the fiber piezo stretcher (FPS). After this adjustment the light intensities are measured again. A histogram of the measured phases during the experiments relative to the setpoints is plotted in Fig.~4b. We note that one could also measure the spin-photon correlations at the second output of the interferometer, which for symmetric states as Eq.1 would yield the same correlations but with opposite sign; however, in the current experiment the slow ($\approx 1$~s) recovery of the detector after being blinded due to metrology light leakage through this output port prevented us from using the second output.
\begin{figure}
	\centering
	\includegraphics[width = 0.4\textwidth]{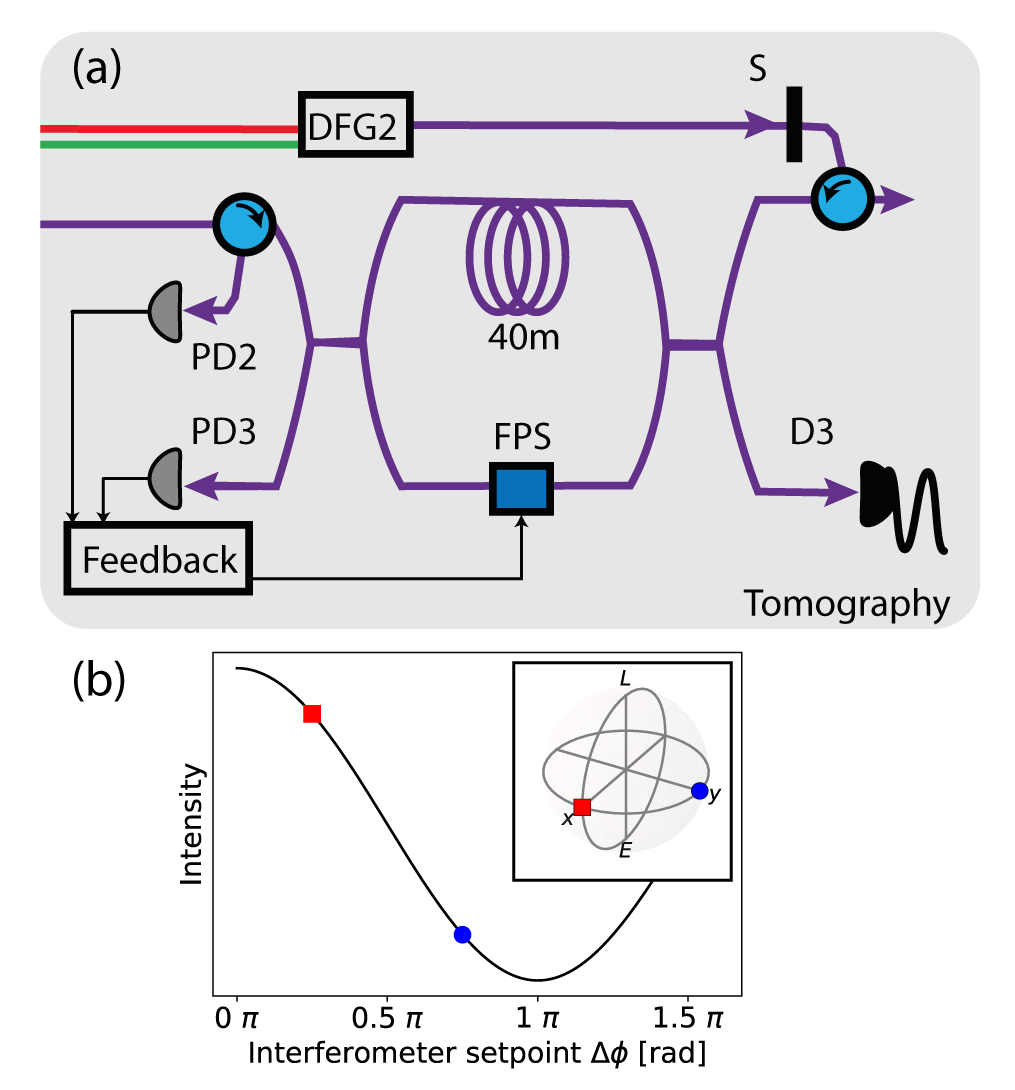}
	\caption{ (a) Polarization-maintaining fiber-based imbalanced interferometer used for the photon state readout in X and Y bases. The frequency-converted single photons are directed into the interferometer. One output port of the interferometer is connected to a superconducting nanowire detector (detector D3). Every second the phase of the interferometer is stabilized. Classical frequency-converted light created by a second DFG setup (DFG2) is sent into the interferometer via a shutter S and a circulator. Light intensities measured by photodiodes PD2 and PD3 are used to generate a feedback signal to the fiber piezo stretcher (FPS) to maintain the target phase $\Delta \phi$. 
	(b) Bloch sphere with the selected photon qubit readout bases indicated on it, and the corresponding phase set points of the imbalanced interferometer.}
\end{figure}
In the remaining 900 ms of each cycle spin-photon correlations are measured using the same protocol as for the ZZ basis (see Fig. 2b). To read out the NV spin state in the appropriate rotated basis, the eigenstates $\ket{\text{X}}$ ($\ket{\text{Y}}$) and the $\ket{\text{-X}}$ ($\ket{\text{-Y}}$) are mapped onto the $\ket{0}$ and $\ket{1}$  states, respectively, by applying an appropriate MW pulse before optical readout.

Figure 4c shows the measured spin-photon correlations in the X and Y basis (bottom), along with expected correlations for the ideal state (top). The letters indicate the spin and photon bases respectively, for example -XX indicates that the NV spin is measured along the -X axis on the Bloch sphere, while the photon is projected on +X. The measured contrast between the correlations and anti-correlations in the X basis is $E_{X} = 0.52 \pm 0.07$ and $E_{Y} = 0.69 \pm 0.07$ in the Y basis.


All data show clear (anti-)correlation between the NV spin qubit and the telecom photonic qubit. With the contrast data from all three orthogonal photon readout bases, we calculate the fidelity $\mathcal{F}$ of our produced state (conditioned on photon detection) to the maximally entangled state of Eq.~1 as
\begin{equation}
\mathcal{F}= \frac{1}{4}\left(1+ E_{X} + E_{Y} + E_{Z}\right),
\end{equation}
yielding a fidelity of $\mathcal{F} = 0.77 \pm 0.03$. This value exceeds the classical boundary of $0.5$ by more than eight standard deviations, proving the generation of entanglement between the NV spin qubit and the frequency-converted photonic qubit. For comparison, reported fidelities for unconverted NV spin-photon entangled states range from $\approx 0.7$~\cite{Togan2010, Trupke2018} to more than $ 0.9$
(estimated from an observed spin-spin entangled state fidelity of $\approx 0.9$~\cite{Hensen.Bell.test}).

The observed fidelity is reduced compared to the ideal value of 1 due to several factors. First, the initial spin-photon entangled state has imperfections, for instance due to photon emission and re-excitation of the NV center during the optical $\pi$-pulse~\cite{Humphreys2018} and small frequency shifts due to spectral diffusion. In addition, the remaining frequency variations of the two locked lasers ($\sim$200~kHz) leads to phase uncertainty between the two terms in Eq.1. All these effects reduce the contrast of the XX and YY correlations, but not that of the ZZ correlations. Second, spontaneous parametric downconversion (SPDC) and Raman scattered photons, produced during the frequency conversion process, add noise to the state as described above and reduce correlations in all bases. Based on these factors, we expect a state fidelity in the range $0.82-0.87$.

\begin{figure}[tb]
	\centering
	\includegraphics[width = 0.4\textwidth]{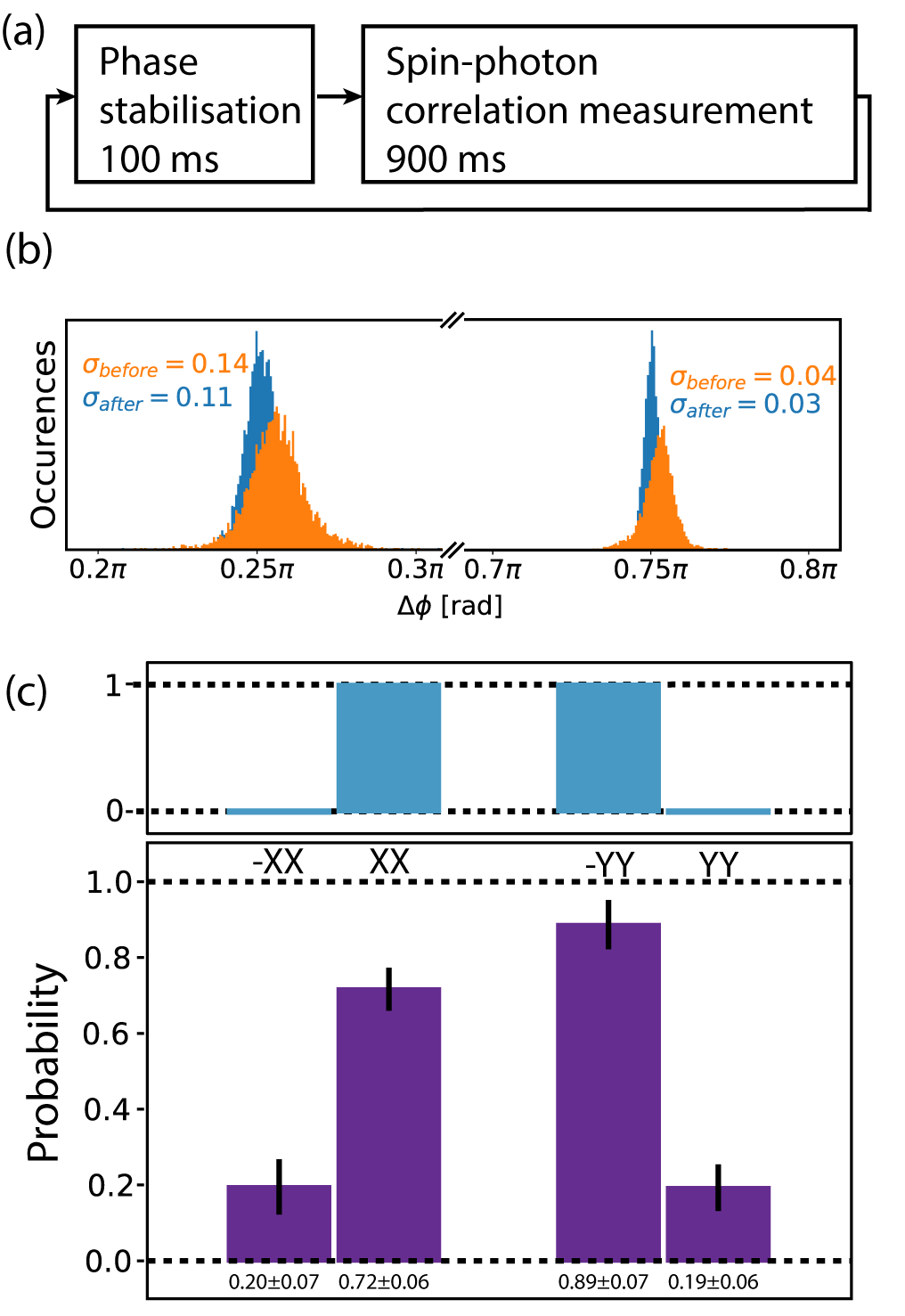}
	\caption{(a) Experimental protocol for measurements in the photon X and Y basis. 
	(b) Measured phase difference $\Delta\phi$ just before stabilization (orange, with 900 ms free evoluation time) and directly after stabilization (blue) for the two setpoints $\Delta\phi_X = \pi/4$ and $\Delta\phi_Y = 3\pi/4$. From the standard deviations in these data, we estimate a residual phase drift of $0.05$ and $0.01$ rad/s for the X and Y photon qubit readout bases, respectively. (c) Results for the correlations in the X and Y basis in purple. The top-panel shows ideal correlations. In total we have measured 1595 photon detection events.}
\end{figure} 

The slight difference between the expected and measured state fidelity could be due to the inaccuracies and fluctuations in setting the interferometer phase setpoint. 
Imperfect interferometer settings result in measurement bases that slightly deviate from the expected X and Y bases, reducing the maximally observable correlations. 
Therefore, the obtained $\mathcal{F}\geq 0.77 \pm 0.03$ sets a lower bound on the true entangled state fidelity.

In conclusion, we demonstrated entanglement between an NV center spin qubit and a time-bin encoded photonic qubit at telecom wavelength, which is an essential step towards long-distance quantum networks based on remote entanglement between NV center nodes. In future experiments the observed state fidelity can be further increased in several ways. A more narrow band frequency filtering after the DFG1 setup would reduce the added noise in the frequency conversion, as the current narrow-band filter has a linewidth $\sim 10$ times larger than the NV-emitted resonant ZPL photons. The signal could be increased by improving the conversion efficiency. Finally, the emission rate of resonant photons and the collection efficiency can be increased by placing the NV center in an optical cavity \cite{Faraon2012, Johnson2015, Hunger2016, Riedel2017, Bogdanovic2017,Englund2018}.


{\it Acknowledgements.} 
We thank L. Schriek, E. Nieuwkoop, J. Lugtenburg and W. Peterse for experimental assistance, and M.~J.~A.~de~Dood and C. Osorio Tamayo for useful discussions. We acknowledge financial support from the Netherlands Organisation for Scientific Research (NWO) through a VICI grant and the Zwaartekracht Grant Quantum Software Consortium and the European Research Council through an ERC Consolidator Grant.\\

\end{document}